\documentclass[12pt]{article}       
\newcommand{\F}{\mathcal{F}}
\newcommand{\R}{\mathcal{R}}

\newcommand{\reals}{{\textrm{I\hspace{-0.5mm}R}}}
\newcommand{\pc}{{\textrm{I\hspace{-0.5mm}P}}}

\begin{document}
%
%
\begin{flushright}DAMTP-2003-110\end{flushright}
\begin{center}
{\Large \bf Almost product manifolds as the low energy\\[5pt] \bf geometry of Dirichlet branes}\\[12pt]
{Frederic P. Schuller}\\[5pt]
{\sl Department of Applied Mathematics and Theoretical Physics,\\ 
University of Cambridge, Cambridge CB3 0WA, United Kingdom\\ 
and\\ 
Perimeter Institute for Theoretical Physics, Waterloo N2J 2W9, Canada; \\
E-mail: fschuller@perimeterinstitute.ca}
\end{center}


 {\small Any candidate theory of quantum gravity must address the breakdown of the classical smooth manifold picture of space-time at distances comparable to the Planck length. String theory, in contrast, is formulated on conventional space-time. However, we show that in the low energy limit, the dynamics of generally curved Dirichlet $p$-branes possess an extended local isometry group, which can be absorbed into the brane geometry as an almost product structure. The induced kinematics encode two invariant scales, namely a minimal length and a maximal speed, without breaking general covariance. Quantum gravity effects on D-branes at low energy are then seen to manifest themselves by the kinematical effects of a maximal acceleration. 
Experimental and theoretical implications of such new kinematics are easily derived. We comment on consequences for brane world phenomenology.}\\[24pt]
{\small {\sl Journal References:} invited article in European Physical Journal C, reprinted in Proceedings of the 41st International School on Subnuclear Physics 2003, Erice, Italy}
%
%
\section{Introduction}
The formulation of a relativistic theory of quantum gravity is one
of the key open questions in fundamental physical theory today.
Attempts to reconcile the principles of general relativity and
quantum theory by employing otherwise tried-and-tested methods
indeed face severe difficulties, suggesting that a considerable
departure from the standard space-time picture may be inevitable.
This becomes dramatically clear in a simple gedankenexperiment.
Assume we want to probe the space-time structure down to
arbitrarily small distances. The position-momentum uncertainty
relation predicts that this can only be done at the cost of
increasingly large fluctuations of the energy-momentum tensor.
These directly translate into fluctuations of the geometry via
Einstein's field equations. A simple calculation shows that if one
aims at resolving lengths on the order of the Planck length
$\ell_P=\sqrt{\hbar G/c^3}$, the described mechanism significantly
disturbs the very space-time distance that one attempts to
resolve. Any candidate theory of quantum gravity is therefore
expected to conceptually involve a fundamental length scale of
order $\ell_P$. Due to the r\^ole of $\ell_P$ as the minimum
resolvable length in the above reasoning, it indeed appears
appropriate to look for a space-time structure in which the Planck
length joins the speed of light as a geometrical invariant. Thus,
it seems inevitable that quantum gravity should be based on such new
kinematics with two invariant scales, at least in the low energy limit.

The conceptual basis of string theory is apparently in direct
opposition to the above reasoning. Standard semi-Riemannian
space-time is upheld as the geometry underlying the formulation of
the theory. A length scale, however, is subtly implemented in the dynamics. 
The fundamental objects of the classical theory are
assumed to be strings of characteristic length $\ell$. A
geometrical action proportional to the world-sheet surface area swept
out by the string in space-time then features the length scale as
an overall factor for dimensional reasons. However, the simple
geometrical formulation of the theory comes at a price. The
corresponding quantum theory is only consistent in 26-dimensional
space-time for the bosonic string, or 10 dimensions if one
includes fermions and supersymmetry. This result challenges phenomenological models
to provide a compelling reasoning of how the observable
4-dimensional universe is supposed to emerge from such a picture.
Proposals to resolve this question have experienced valuable new
input from the discovery of Dirichlet-$p$-branes \cite{Polchinski} as
non-perturbative solutions in string theory. D$p$-branes are
$(p+1)$-dimensional sub-manifolds of the 10- or 26-dimensional string
target space, defined by the property that open strings can end on
them. Their phenomenological significance arises from ideas to
devise models of the observable universe as a D$3$-brane
propagating in the higher dimensional space-time \cite{Randall}. The dynamics of
(and the new physics seen on) such brane-worlds originate from the
interaction with the strings propagating in the higher dimensional
space-time. It is therefore of utmost interest to study and understand the properties of D-branes, as they constitute the fundamental building blocks of all such brane world theories. Properties of these building blocks have an impact on any phenomenological scenario in which they are involved, and are hence largely model-independent.  \\ 

The key observation of this paper is that the low energy dynamics
of D$p$-branes possess a hidden invariance, which can be absorbed
into the world-volume geometry of the brane. This geometry can be
viewed, alternatively, either as a module bundle over a
semi-Riemannian manifold \cite{FPSHP}, or as a tangent bundle of the standard
space-time sub-manifold with an almost product structure. New
geometry is synonymous with new kinematics \cite{BIK}. In the present case, the local Lorentz symmetry
remains intact, but is extended to a larger local gauge group, by
transformations to rotating and sub-maximally accelerated frames, with
the maximum acceleration given by the inverse $1/\ell$ of the
string length scale. In other words, the original {\sl dynamical} encoding of a
length scale in string theory finally implies a particular
{\sl kinematical} implementation on $(p+1)$-dimensional manifolds, as it is expected of a candidate theory of quantum gravity.


\section{Pseudo-complex module bundles over D$p$-branes}
The exact dynamics of D$p$-branes is determined by the interaction of
the open strings ending on the brane with any other strings in the
theory. The analysis of the resulting dynamics is, of course,
prohibitively difficult. In the low energy limit, however, only
the massless string modes contribute. For technical simplicity, we consider in this article the toy model of bosonic string theory. The quantum spectrum of bosonic string theory contains a tachyon, which renders Dirichlet branes unstable. We will ignore this issue altogether, but remark that the following developments can be extended in a straightforward manner to type I superstring theory. The irreducible components
of the second rank tensor modes of massless closed bosonic strings give
rise to effective background fields $G_{(MN)}$, $B_{[MN]}$ and
$\Phi$, where $M,N=0, \dots, 25$. These can be identified as the
classical target space metric, Neveu-Schwarz two-form potential,
and the dilaton. The massless vector modes of open strings ending
on the D$p$-brane produce an effective gauge field $A_\mu$, where
$\mu=0, \dots, p$. It has been shown that in the low energy limit,
the dynamics of this gauge field $A$ is given by the
Dirac--Born--Infeld (DBI) action \cite{Leigh}
\begin{equation}\label{DBIorig}
  \int_{\textrm{brane}} \sqrt{\det(g_{\mu\nu})} dx^0\wedge\dots\wedge dx^p  e^{-\Phi} \sqrt{\det({\delta^\mu}_\nu + {B^\mu}_\nu + \ell^2 {F^\mu}_\nu)},
\end{equation}
where $F_{\mu\nu} = \partial_{[\mu} A_{\nu]}$, and $g_{\mu\nu}$
and $B_{\mu\nu}$ are the pull-backs to the brane of the target
space fields $g_{MN}$ and $B_{MN}$, respectively. Indices $\mu,
\nu$ are lifted and lowered using the induced metric on the brane.
Note that for $p=3$ the action (\ref{DBIorig}) can be viewed as a theory of non-linear electrodynamics \cite{GWG}, and has in fact been devised as such in the 1930s \cite{BornInfeld} in order to covariantly regularize the energy divergence of the electrostatic field of a charged point particle in Maxwell theory. 
Expansion of the determinant, using the identity $\det(1+\F) =
\exp \textrm{tr} \ln(1 + \F)$, with  $\F = B + \ell^2 F$, shows that only even powers of $\F$
contribute to the action. We
can hence multiply $\F$ by a number $I$ satisfying $I^2=+1$,
without changing the action (\ref{DBIorig}) at all.
In order to clear up the notation, we use the shorthand $\omega =
\sqrt{det(g_{\mu\nu})} dx^0\wedge\dots\wedge dx^p$ for the volume
form on the brane, so that we have
\begin{equation}\label{DBI}
  \int_{\textrm{brane}} \omega  e^{-\Phi}\sqrt{\det({\delta^\mu}_\nu + {\F^\mu}_\nu)} = \int_{\textrm{brane}} \omega  e^{-\Phi}\sqrt{\det({\delta^\mu}_\nu + I {\F^\mu}_\nu)}.
\end{equation}
Note that
$\F_{\mu\nu}$ are the components of a $(0,2)$-tensor field on the
brane $\Sigma$, i.e., $\F(p)$ is a linear map $T_p\Sigma \otimes
T_p\Sigma \longrightarrow \reals$ for any point $p\in \Sigma$. The
addition on the left hand side of Eq. (\ref{DBI}) is the addition
in the vector space of real $(0,2)$ tensors. If we assume that
$I\in\reals$ (i.e. $I=\pm 1$) then the addition on the right hand
side is the same, and also well-defined. It will turn out to be
enlightening, however, to take $I\not\in\reals$ and to define an
algebraic extension
\begin{equation}
  \pc := \{a + I b | a, b \in \reals\}
\end{equation}
of $\reals$, where we identify $\reals \equiv \{a + I b | a \in
\reals, b=0 \}$. The set $\pc$ equipped with the addition and
multiplication inherited from $\reals$ fails to be a field due to
the existence of zero-divisors of the form $\lambda(1\pm I)$,
where $\lambda\in\reals$, which do not possess multiplicative
inverses. However, $(\pc,+,\cdot)$, to whose elements we will
refer as pseudo-complex numbers, is a commutative ring. 
One also finds the terms double numbers, hyperbolic complex or para-complex numbers for the ring $\pc$ in the literature, indicating that this simple structure has ample applications \cite{Hucks}, but is little known and hence re-invented time and again.   
The commutativity of the ring $\pc$
allows, in particular, for the construction of Lie algebras over
$\pc$. Note that in accordance with the mathematical literature,
vector space like structures over a ring $R$ rather than a number field will be
called $R$-modules in this paper. Taking $I\not\in\reals$ therefore
enforces the pseudo-complexification $(T_p\Sigma)_\pc := \{v + I
w\, |\, v, w \in T_p\Sigma \}$ of the real tangent spaces, such
that tensors of type $(r,s)$ are now $\reals$-linear maps
\begin{equation}
  \bigotimes_{s} (T_p\Sigma)_\pc \longrightarrow \bigotimes_r
  (T_p\Sigma)_\pc,
\end{equation}
constituting a real vector space for each pair $(r,s)$. This renders
the addition on the right hand side of Eq. (\ref{DBI})
well-defined if $I\not\in\reals$. In summary, the insertion of the
pseudo-imaginary unit $I$ into the DBI action is valid if one
extends the tangent bundle $T\Sigma$ of the brane world-volume to
a $\pc$-module bundle over $\Sigma$ with typical fiber
$\pc^{p+1}$. In like fashion, the frame bundle $L(\Sigma)$ of the real manifold $\Sigma$ is replaced by the pseudo-complexified frame bundle $L_\pc(\Sigma)$. 
In general relativity, an observer is given by a curve $e: \reals \longrightarrow L(\Sigma)$ in the frame bundle, where the metric $g$ on $\Sigma$ is used to orthonormalize the frame such that $g(e_a,e_b)=\eta_{ab}$, where $a,b= 0, \dots, p$ and $\eta$ is the Minkowski metric. The local $O(1,p)$ gauge group of the Lorentzian manifold $\Sigma$ parameterizes the freedom to choose equivalent orthonormal frames. The frame vector $e_0$ is taken to be the unit tangent to the curve $\pi(e)$ on $\Sigma$, where $\pi$ is the canonical bundle projection. The covariant change of the frame along the observer's world-line $\pi(e)$ is then parameterized by an anti-symmetric Lorentz tensor $\Omega_{ab}$, such that 
\begin{equation}
  \nabla_{e_0} e_a = {\Omega_a}^b e_b.
\end{equation} 
The translational $p$-acceleration of the observer is $\Omega_{0\alpha}$, with $\alpha=1, \dots, p$. With respect to an observer whose spatial frame vectors $e_\alpha$ are parallely transported along $\pi(e)$, our observer $e$ possess angular velocity $\Omega_{\alpha\beta}$ in the $\alpha\beta$-plane \cite{MTW}. 

The Frenet-Serret tensor $\Omega$ is therefore implicitly contained in the choice of any given observer $e$.
 When extending the real frame bundle $L(\Sigma)$ to the pseudo-complexified one $L_\pc(\Sigma)$, we choose to explicitly encode the Frenet-Serret tensor in the pseudo-imaginary part of a pseudo-complex frame $E: \reals \longrightarrow L_\pc(\Sigma)$,
\begin{equation}\label{Edef}
  E_a := {\gamma_a}^b (\delta_b^c + I \ell {\Omega_b}^c) e_c,
\end{equation}
where we have included the length scale $\ell$ for later interpretational convenience.
The overall tensor factor $\gamma$ is a normalization factor such that
\begin{equation}
  g(E_a,E_b) = \eta_{ab}.
\end{equation}  
The freedom of choice for such frames is now obviously parameterized by the gauge group $O_\pc(1,p) \cong O(1,p) \times O(1,p)$. This decomposition of the pseudo-complex Lorentz group $O_\pc(1,p)$ into two copies of the real Lorentz group is easily seen in the zero-divisor decomposition of $\pc$. The real Lorentz group presents a proper subgroup of $O_\pc(1,p)$, and is diagonally embedded in this decomposition. Note that this means that the standard local Lorentz symmetry is fully preserved by the pseudo-complexification of the frame bundle. Thus we can identify the inertial frames of general relativity with those of the pseudo-complexified theory, which allows to maintain the strong equivalence principle. 

In order to exhibit the physical interpretation of the action of $O_\pc(1,p)$, first note the following polarization formula. Any $\Lambda \in O_\pc(1,p)$ can be written as a unique product of a real Lorentz transformation $L \in O(1,p)$ and a pseudo-complex Lorentz transformation $K$ with purely pseudo-imaginary coefficients,
\begin{equation}
  {\Lambda^a}_b = {K^a}_m {L^m}_b,
\end{equation}
as can be easily shown in the zero-divisor decomposition. As the action of real Lorentz transformations $L$ is well-understood, the polarization formula allows to analyze the meaning of general $O_\pc(1,p)$ transformations by study of transformations of type $K=\exp(\omega_{mn} I M^{mn})$, i.e., transformations with purely pseudo-imaginary coefficients. 
Consider an unaccelerated and non-rotating observer at time $\tau$, so that $\Omega(\tau)=0$. The pseudo-complex frame at this instant is then simply $E_a = e_a$. A real Lorentz transformation $L$ will simply re-define the real frame and, of course, map the real Lorentz tensor $\Omega=0$ onto itself. A pseudo-complex Lorentz transformation of type $K$, however, will yield a transformed frame 
\begin{equation}\label{transE}
  E = \cosh(\omega_{mn}M^{mn}) \left(1 + I \tanh(\omega_{mn}M^{mn})\right) e,
\end{equation}  
corresponding to mapping the Frenet-Serret tensor $\Omega=0$ to
\begin{equation}\label{transOmega}
  \Omega \longrightarrow \ell^{-1} \tanh(\omega_{mn}M^{mn}).
\end{equation}
This corresponds to a transformation to a non-inertial frame, with the values for the $p$-acceleration and the angular velocities to be read off from the corresponding components of $\Omega$. In the following, we will only consider pseudo-complex frames that are locally continuously connected to inertial frames, i.e., frames $E_a$ such that $E_a = {\Lambda^m}_a e_m$ for some real frame $e$, and $\Lambda$ an element of the connection component of the identity of the pseudo-complex Lorentz group. We call such frames admissible.
Now consider the phenomenologically interesting case $p=3$. There are two real Lorentz invariants encoded in the Frenet-Serret tensor,
\begin{eqnarray}
  I_1 &=& \frac{1}{2}\Omega_{ab} \Omega^{ab} = \mathbf{a}^2 - \mathbf{L}^2,\\
  I_2 &=& \frac{1}{2}\Omega_{ab} (*\Omega)^{ab} = 2 \mathbf{a}.\mathbf{L},
\end{eqnarray}
where $*\Omega$ denotes the Hodge dual of the two-form $\Omega$. The 3-vectors $\mathbf{a}$ and $\mathbf{L}$ are the translational acceleration and angular velocity of the observer.  
Restricting attention to observers with admissible frames, we can always apply local $O^e_\pc(1,3)$ transformations to obtain a Fermi-Walker transported observer, i.e., $\mathbf{L}=\mathbf{0}$. For such non-rotating observers, it is easy to see by direct calculation that the condition that the frame be admissible corresponds to requiring a covariant upper limit on scalar accelerations,
\begin{equation}
  \mathbf{a}^2 < 1/\ell^2,
\end{equation}
given by the inverse of the string length scale. The fact that a maximal acceleration arises as a consequence of a minimal length scale, through the above relation, is not too surprising. Consider the following simple causality argument: A relativistic observer of scalar acceleration $g$ cannot set up, in an operationally well-defined manner, a coordinate system that extends more than a distance $1/g$ in any spatial direction, because the Rindler horizon makes him causally disconnected from certain regions of space-time. Now if there is a spatially extended object of characteristic minimal length $\ell$, then uniform acceleration of the whole object, at a value larger than $1/\ell$, would causally disconnect parts of this object. 

Before pressing on with the application to the D$p$-brane geometry, we list the irreducible second rank tensor representations of the pseudo-complex Lorentz group. Of immediate physical interest is the connection component of the identity of the pseudo-complexified Lorentz group, which we denote by $O_\pc^e(1,p)$. Any element $\Lambda$ of the defining vector representation $\R_v$ of $O^e_\pc(1,p)$ can be generated by exponentiation of the standard Lorentz generators ${(M^{mn})^a}_b = \eta^{m a} \delta^n_b - \eta^{n a} \delta^m_b$ with pseudo-complex parameters $\omega_{mn} \in \pc$,
\begin{equation}
  \Lambda(\omega) = \exp(\omega_{mn}M^{mn}), \qquad m, n = 0, \dots, p. 
\end{equation}
It can be shown \cite{FPSHP} that the pseudo-complex conjugate representation $\R_v^*$ is equivalent to $\R_v$ over $\reals$, but inequivalent over $\pc$. All irreducible second rank tensors are therefore contained in 
\begin{equation}
  \R_v \otimes_\reals \R_v \cong \R_t \oplus \R_a \oplus \R_s \oplus \R_H \oplus \R_{\overline{H}},
\end{equation}
if we use a tensor product over $\reals$, or
\begin{eqnarray}
  \R_v \otimes_\pc \R_v &\cong& \R_t \oplus \R_a \oplus \R_s,\\
  \R_v \otimes_\pc \R_v^* &\cong& \R_H \oplus \R_{\overline H},\label{RRs}
\end{eqnarray}
if we use a tensor product over $\pc$. Elements of the representation spaces $\R_t$, $\R_a$, $\R_s$, $\R_H$, $\R_{\overline H}$ can be concisely encoded in a pseudo-complex trace, and pseudo-complex symmetric, anti-symmetric, hermitian or anti-hermitian matrices. Both methods to take tensor products yield the same irreducible representations finally. We mention both in order to illustrate that care must be taken in specifying whether one deals with $\reals$-linear or $\pc$-linear structures. From the tensor product (\ref{RRs}), we read off the transformation behavior of hermitian second rank tensors $H \in \R_H$ as
\begin{equation}
  H_{ab} \longrightarrow {\Lambda^m}_a {\Lambda^{* n}}_b H_{mn} 
\end{equation}
under $O_\pc(1,p)$-transformations, where $\Lambda \in \R_v$. Rewriting the Dirac--Born--Infeld action in terms of the pseudo-hermitian tensor $H$, whose local frame components are given by${H^a}_b := E^a_\mu E^{* \nu}_b({\delta^\mu}_\nu + I {\F^\mu}_\nu)$, yields
\begin{equation}\label{DBI2}
  \int_{\Sigma} \omega e^{-\Phi}\sqrt{\det(E^\mu_a E^{*b}_\nu {H^a}_b)}.
\end{equation}
This expression is manifestly invariant under $O_\pc(1,p)$ transformations. Let us briefly summarize what we have achieved by casting the Dirac--Born--Infeld action into the form (\ref{DBI2}). In its original form (\ref{DBIorig}), the length scale $\ell$ appears as a numerical constant, without any geometrical meaning. Extending the real frame bundle of the brane to its pseudo-complexified version allows to re-write the DBI action in the fully equivalent form (\ref{DBI2}). The length scale $\ell$, however, now appears as an invariant of the orthogonal group $O^e_\pc(1,p)$, formally on an equal footing with the invariant speed of light. The kinematical interpretation of pseudo-complex Lorentz transformations identifies $1/\ell$ as the maximum admissible (Lorentz-)scalar acceleration for a non-rotating observer. For a rotating observer with angular velocity $\mathbf{L}$, the maximal acceleration is shifted up to $\sqrt{\ell^{-2} + \mathbf{L}^2}$. We have thus achieved our goal of geometrizing the length scale $\ell$, in a way that is consistent with the low energy dynamics of Dirichlet $p$-branes.

\section{Almost product manifolds}
So far, we considered a semi-Riemannian manifold $\Sigma$ with pseudo-complexified tangent spaces. From a mathematical point of view, this is a somewhat hybrid structure, and therefore leads us to the natural question of whether this pseudo-complex structure of the tangent spaces can be absorbed into the manifold structure itself. The purpose of the present section is to cast this question into a precise form, and to find to which extent such a reformulation is possible. From a physical point of view, we might want to restrict our attention to Fermi-Walker transported observers, as we can always arrange for such systems experimentally by means of gyroscopes. Consider the frame vector $E_0$ of an observer on a low energy D$p$-brane with $\Omega_{\alpha\beta}=0$ for $\alpha, \beta = 1, \dots, p$. Using Eq. (\ref{Edef}), we find an expression for this frame vector in terms of the $(p+1)$-velocity $u$ and covariant acceleration $a = \nabla_u u$ of the observer's world-line $x$,
\begin{equation}
  E_0 = \frac{u + I \ell a}{\sqrt{1-\ell^2 a^2}}.
\end{equation}
Representing the unit $1$ and pseudo-imaginary unit $I$ in $\pc=\reals\oplus\reals$ by matrices
\begin{equation}
  1 = \left(\begin{array}{cc}1 & 0 \\ 0 & 1\end{array}\right),\qquad I = \left(\begin{array}{cc}0 & 1 \\ 1 & 0\end{array}\right),
\end{equation}
and identifying the pseudo-complex module $\pc^n$ with $\reals^n \oplus \reals^n$, the normalization condition $g(E_0,E_0)=1$ can be written as the two conditions
\begin{eqnarray}
  \gamma^2 (g\otimes 1)\left(u \oplus \ell a, u \oplus \ell a\right) &=& 1,\label{tr1}\\
  \gamma^2 (g\otimes I)\left(u \oplus \ell a, u \oplus \ell a\right) &=& 0,\label{tr2}
\end{eqnarray}
where $\gamma = 1/\sqrt{1 - \ell^2 a^2}$.
Now consider the natural lift of the curve $x$ in $\Sigma$ to the curve $X = x \oplus \ell u$ on the tangent bundle $T\Sigma$. If $\tau$ is the natural parameter of the curve $x$ with respect to the metric $g$ on $\Sigma$, then define the new parameter $\omega = \tau/\gamma$, which we will soon identify as the natural parameter of the lifted curve $X$ with respect to a particular metric on $T\Sigma$. It is easily shown that in terms of the lifted curve $X$, the normalization conditions read
\begin{eqnarray}
  g^D(\frac{dX}{d\omega},\frac{dX}{d\omega}) &=& 1, \label{gdfinal}\\
  g^H(\frac{dX}{d\omega},\frac{dX}{d\omega}) &=& 0, \label{ghfinal}
\end{eqnarray} 
where $g^D$ and $g^H$ are the so-called diagonal and horizontal lifts of the space-time metric $g$ to the tangent bundle \cite{YanoIshihara}. Note that $dX/d\omega = \gamma(u \oplus \ell du/d\tau)$ is not identical to $E_0 = \gamma(u\oplus \ell \nabla_u u)$. However, the connection coefficients in $\nabla_u u$ are absorbed into the definition of $g^D$ and $g^H$, whose components in the induced frame on $T\Sigma$ can be easily derived from the stated equivalence of (\ref{tr1}-\ref{tr2}) to (\ref{gdfinal}-\ref{ghfinal}). As is shown in differential geometry, both $g^D$ and $g^H$ are globally defined semi-Riemannian metrics on $T\Sigma$. From Eq. (\ref{gdfinal}), we see that the parameter $\omega$ is the natural parameter of the curve $X$ with respect to the metric $g^D$, as anticipated above. 
 We hence obtain a bi-metric tangent bundle picture $(T\Sigma, g^D, g^H)$ for the Dirichlet brane geometry, equivalent to the module-bundle approach for non-rotating observers.

The question of whether the pseudo-complex structure of the tangent spaces ultimately originates from a manifold with pseudo-complex coordinates, can now be rigorously addressed. The bi-metric structure $(T\Sigma, g^D, g^H)$ can be reconstructed from a metric tangent bundle 
\begin{equation}\label{amp}
  (T\Sigma, g^D, F)
\end{equation}
with a globally defined almost product structure $F := {(g^D)}^{-1} g^H$. An almost product structure is a $(1,1)$ tensor $F$, such that $F^2$ is the identity transformation on the tangent spaces $T_QT\Sigma$ of the tangent bundle $T\Sigma$ for all $Q\in T\Sigma$. Product and almost product manifolds have been explored in the mathematical literature, and there exist integrability theorems analogous to those for complex and almost complex manifolds. In particular, the vanishing of the Nijenhuis tensor 
\begin{equation}
  {N_{LM}}^J := (\partial_K{F^J}_L - \partial_L{F^J}_K){F^K}_M - (L \leftrightarrow M), \qquad J, K, L, M = 0, \dots, 2p+1
\end{equation} 
is necessary and sufficient \cite{YanoCM} for the almost product structure to be induced from a manifold with pseudo-complex local charts $\pc^{p+1}$. However, for the almost product structure at hand, $F= (g^D)^{-1} g^H$, the Nijenhuis tensor can be calculated explicitly and is seen to vanish if and only if the base manifold $\Sigma$ is flat. This is, of course, not the generic case for a Dirichlet brane. This answers our question, of whether the 
pseudo-complex structure of the tangent spaces can be fully absorbed into pseudo-complex coordinates, to the negative. However, in the tangent bundle formulation, the normalization conditions (\ref{gdfinal}-\ref{ghfinal}) provide a clear physical interpretation for the r\^oles of the metrics $g^D$ and $g^H$. The requirement that tangent bundle curves are null with respect to $g^H$ is simply a reformulation of the orthogonality of covariant velocity and acceleration, $g(u,a)=0$ for any timelike world-line. The normalization of the tangent bundle vector $dX/d\omega$ with respect to $g^D$ is equivalent to requiring that there is an upper bound on admissible covariant accelerations, $g(a,a)<1/\ell^2$. 

The partial geometry $(T\Sigma, g^D)$ has been studied before \cite{Brandt,Caianiello} as a maximal acceleration geometry, but without contact to any well-studied candidate theory of quantum gravity. It is remarkable that the low energy dynamics of Dirichlet branes imply just this geometry, and complete it by requiring that the tangent bundle is further equipped with a second metric $g^H$, or, equivalently, an almost product structure $F$. Indeed, analogy with the symplectic structure of classical phase space would apparently rather suggest almost complex tangent bundles $(T\Sigma, g^D, J)$, with $J^2 = -1$. There is, however, a theorem due to Tachibana and Okumura \cite{Tachibana}, that shows that simultaneous covariant constancy of both structures, $\nabla g^D = 0$ and $\nabla J = 0$    
(which is required if one wants to invoke a strong principle of equivalence), is possible if and only if the base manifold $\Sigma$ is flat. In contrast, there is a connection $\nabla^H$ on $T\Sigma$, the horizontal lift \cite{YanoIshihara} of the Levi-Civita connection on $(\Sigma, g)$, which renders both the metric $g^D$ and the almost product structure $F$ simultaneously covariantly constant.
The structure of low energy Dirichlet branes therefore induces a geometry that is consistent with the strong principle of equivalence. We have seen this compatibility of the maximal acceleration/minimal length geometry with the strong equivalence principle before in the module bundle picture, and thus the automatic circumvention of the Tachibana-Okumura theorem in the tangent bundle approach provides a non-trivial consistency check on that result. 

\section{Applications}
We give two examples for applications of the Dirac--Born--Infeld kinematics, one each for the module bundle picture and almost product manifold picture, respectively.
The Thomas precession of the spatial frame of an observer in circular motion with respect to an inertial frame is a standard result in special relativity. It is brought about essentially by the structure of the real Lorentz algebra, in particular the commuting of two independent boost generators up to a rotation generator,
\begin{equation}
  [M^{0\alpha},M^{0\beta}] = c^{-2} M^{\alpha\beta}.
\end{equation}   
As the orbiting observer has non-constant velocity, one must perform successive infinitesimal Lorentz boosts, in order to analyze the parallel transport of the spatial frames attached to the observer, using the above commutation relations. In the non-relativistic limit $c\rightarrow\infty$, the effect vanishes. 
Such an observer in circular motion also undergoes a non-constant acceleration. In the presence of a length scale $\ell$, changes to accelerating frames are generated by $I M^{o\alpha}$, as we saw from Eq. (\ref{transOmega}). Successive infinitesimal transformations of this type effect an additional rotation of the spatial frame, because 
\begin{equation}
   [I M^{0\alpha},I M^{0\beta}] = \ell^2 M^{\alpha\beta}.
\end{equation} 
The corrected Thomas precession rate for an observer performing circular motion of radius $R$ and angular velocity $\omega$ is found \cite{BIKletter} to be
\begin{equation}
\left(\sqrt{(1-R^2\omega^2/c^2)(1-R^2\ell^2\omega^4/c^2)}-1\right)\omega,
\end{equation}
deforming the standard result by the length parameter $\ell$. However, the real Lorentz symmetry algebra $so(1,p)$ is a proper sub-algebra of the pseudo-complex algebra $so_\pc(1,p)$, meaning that Lorentz symmetry is not affected by the presence of the invariant length scale. This shows, as a corollary, that high precision measurements of the Thomas precession in atomic physics cannot possibly falsify Lorentz symmetry, as is often assumed. Such experiments rather yield a lower bound on the hypothetical maximal acceleration. 

The second example is more of theoretical interest, and uses the above result that in the case of a flat space-time $\Sigma$, the almost product structure on $T\Sigma$ is integrable, i.e., can be absorbed into pseudo-complex coordinates. A careful study \cite{FPSCQG} of free quantum field theory on pseudo-complexified Minkowski space $\pc^{p+1}$ shows that the propagators of any tensor field are automatically Pauli-Villars regularized with the regularization parameter given by the inverse length scale $1/\ell$. From a representation theoretical point of view, this result is understood from the fact that an irreducible representation of the pseudo-complexified Poincar\'e group (being the isometry group on $\pc^{p+1}$) accommodates two irreducible representations of the real Poincar\'e group, of equal spin (helicity) but generically different mass. Invoking a correspondence principle to standard quantum field theory in the limit $\ell \rightarrow 0$, one then observes that one of these real particles acts as a Pauli-Villars regulating Weyl ghost of the other, proper, particle. Taking the standard relativistic limit $\ell\rightarrow 0$ after, rather than before, the construction of a quantum field theory therefore corresponds exactly to the Pauli-Villars regularization prescription. The isometry group $O_\pc(1,p)$ apparently captures the regularization of the classical Dirac--Born--Infeld dynamics in a kinematical way.

\section{Conclusion}
Starting from the Dirac--Born--Infeld action as the low energy dynamics of a gauge field $A$ on a Dirichlet $p$-brane in bosonic string theory, we found that the length scale appearing in the fundamental string dynamics finally gives rise to relativistic kinematics on the brane that preserve this length scale as a geometrical invariant. This finding implies, independently of string theory, two equivalent techniques for the extension of Lorentzian manifolds such as to encode a length scale as a geometrical invariant. The first technique consists in the pseudo-complexification of the individual tangent spaces of the space-time manifold. The resulting module bundle structure is particularly well-suited for the discussion of observers in the new geometry. These still enjoy local Lorentz symmetry, but the extended local isometry group further contains transformations to rotating and accelerated frames. For non-rotating observers, however, only frames of sub-maximal acceleration are continuously connected to inertial frames. The maximal acceleration scale is given by the inverse length scale that originally entered the string action. A surprising immediate consequence of these minimal length/maximal acceleration kinematics is a correction to the Thomas precession. The module bundle formalism naturally allows for the discussion of rotating coordinate systems, but as we can always arrange for Fermi-Walker transported spatial frames by the use of gyroscopes, one can focus one's attention to non-rotating observers. Such a restriction permits to cast the new geometry into the form of a metric structure on the space-time tangent bundle, additionally equipped with a particular almost product structure. Both the tangent bundle metric and the almost product structure are globally defined lifts of the space-time metric. This lift of the metric structure to the tangent bundle presents the second technique for a geometrical implementation of a fundamental length scale, applicable to any Lorentzian manifold. 
The tangent bundle picture is particularly adapted to answer geometrical questions about the theory, as the whole apparatus of differential geometry on the tangent bundle is available. In particular, the theory of product manifolds shows that the pseudo-complex structure on the tangent spaces does not derive from a manifold with pseudo-complex coordinates in the presence of gravity. In the absence of gravity, however, the kinematical group reduces to the isometry group of pseudo-complexified space-time, allowing for the definition of sub-maximally accelerated quantum particles as irreducible representations of the pseudo-complexified Poincar\'e group. In this setting, a conjecture \cite{Nest} on the regularizing effect of a maximal acceleration in quantum field theory can be made precise and proved \cite{FPSCQG}.

In particular on D3-branes, the exhibited string theoretically induced maximal acceleration kinematics are of direct interest for brane world phenomenology. As is illustrated by the case of the Thomas precession, the mathematically simple structure of the theory allows its ready application. The merit of the presented approach to minimal length kinematics roots in the following facts. Any generically curved Lorentzian manifold can be extended such as to geometrically encode a minimal length scale in a covariant way. The employed mathematics consists of standard techniques in differential geometry and linear algebra. The preservation of the local Lorentz symmetry allows to maintain the strong equivalence principle, while the extended local isometry group makes non-trivial predictions for accelerated observers. Finally, its derivation from the low energy dynamics of Dirichlet branes makes rigorous contact with string theory, with implications for string phenomenology and the prospect of future insights into the presented questions inspired by string theory.
\section*{Acknowledgments}
The author thanks G. 't Hooft for the opportunity to speak in the new talent sessions at the 41st International School on Subnuclear Physics in Erice, Italy,
and the University of Cambridge for financial support.


\begin{thebibliography}{5}

\bibitem{Polchinski}
J.~Polchinski,
Phys.\ Rev.\ Lett.\  {\bf 75}, 4724 (1995)

\bibitem{Randall}
L.~Randall and R.~Sundrum,
Phys.\ Rev.\ Lett.\  {\bf 83}, 4690 (1999)

\bibitem{FPSHP}
F.~P.~Schuller and H.~Pfeiffer,
Phys.\ Lett.\ {\bf B578}, 402-408 (2003)

\bibitem{BIK}
F.~P.~Schuller,
Annals Phys.\  {\bf 299}, 174 (2002)

\bibitem{Leigh}
R.~G.~Leigh,
Mod.\ Phys.\ Lett.\ A {\bf 4}, 2767 (1989).

\bibitem{GWG}
G.~W.~Gibbons and K.~Hashimoto,
JHEP {\bf 0009}, 013 (2000)

\bibitem{BornInfeld}
M.~Born and L.~Infeld,
Proc.\ Roy.\ Soc.\ Lond.\ A {\bf 144}, 425 (1934).

\bibitem{Hucks}
J.~Hucks,
J.\ Math.\ Phys.\  {\bf 34}, 5986 (1993).


\bibitem{MTW} 
C.~W.~Misner, K.~S.~Thorne and J.~A.~Wheeler, {\sl Gravitation}, W.~H.~Freeman 1973

\bibitem{YanoIshihara} K.~Yano and S.~Ishihara, {\sl Tangent and Cotangent Bundles}, Marcel Dekker New York 1973

\bibitem{YanoCM} K.~Yano, {\sl Differential Geometry on complex and almost complex spaces} vol. 49 Macmillan Intl. Series of monographs in pure and appl. math., New York 1965

\bibitem{Brandt}
H.~E.~Brandt,
Found.\ Phys.\ Lett.\  {\bf 2}, 39 (1989).

\bibitem{Caianiello}
E.~R.~Caianiello,
Lett.\ Nuovo Cim.\  {\bf 32}, 65 (1981).

\bibitem{Tachibana} S.~Tachibana and M.~Okumura, Tohoku Math. Jour., {\bf 14}, 156-161 (1962)

\bibitem{BIKletter}
F.~P.~Schuller,
Phys.\ Lett.\ B {\bf 540}, 119 (2002)

\bibitem{Nest}
V.~V.~Nesterenko, A.~Feoli, G.~Lambiase and G.~Scarpetta,
Phys.\ Rev.\ D {\bf 60}, 065001 (1999)

\bibitem{FPSCQG}
F.~P.~Schuller, M.~Wohlfarth and T.~Grimm,
Class.\ Quant.\ Grav.\  {\bf 20}, 4269 (2003)
\end{thebibliography}
\end{document}